\documentclass[journal,12pt,3p]{elsarticle}

\usepackage{amssymb}
\usepackage{comment}
\usepackage{algorithmicx}
\usepackage[utf8]{inputenc}
\usepackage{algorithmicx}
\usepackage{algpseudocode}

\usepackage{multirow}
\usepackage{graphicx}
\usepackage{lineno}
\newcommand{\R}{\mathbb{R}}

\usepackage{amsmath}
\usepackage{tikz,calc}

\newtheorem{proposition}{Proposition}

\journal{Environmental Monitoring and Software} 

\begin{document}
\bibliographystyle{plainnat}
\begin{frontmatter}

\title{A Watershed Delineation Algorithm for 2D Flow Direction Grids}
\tnotetext[label0]{This is only an example}

\author[label1]{Scott Haag}
\address[label1]{Academy of Natural Sciences, Drexel University 1900 Benjamin Franklin Parkway, Philadelphia, PA 19103, USA}


\ead{smh362@drexel.edu}
 
\author[label2]{Ali Shokoufandeh}
\address[label2]{Department of Computer Science, Drexel University, Philadelphia, PA 19105, USA\fnref{label4}}
\ead{ashokouf@cs.drexel.edu}

\begin{abstract}
In this paper we propose an algorithm and associated data model for creating a watershed boundary using a 2D Flow Direction Grid. Flow Direction Grids (FDGs) are common abstractions for hydrodynamic models and are utilized for delineating physical systems (e.g. watersheds, fluvial, and non-fluvial flow paths).  The proposed algorithm and associated data model provides geometric speed increases in watershed boundary retrieval while keeping storage constraints linear in comparison to existing techniques. The algorithm called Haag Shokoufandehs' March (HSM) relies on an existing data structure, the modified nested set model, originally described by Celko and applied to hydrodynamic models by Haag and Shokoufandeh in 2017.  The proposed algorithm creates watershed boundaries by marching around the edges of its' corresponding region, never entering the internal area.  In contrast to existing algorithms that scales in proportional to the area of the underlying region, the complexity of the HSM algorithm is proportional to the boundary length. Results for a group of tested watersheds ($n$ = 14,718) in the $\approx$ 36,000 km$^2$ Delaware River Watershed show a reduction of between 0 and $99\%$ in computational complexity using a 30 m DEM vs. existing techniques. Larger watersheds have a consistent reduction in the number of (read) operation  complexity, with the largest watershed resulting in $\approx$ 35 million reads using traditional techniques compared to $\approx$ 45 thousand using the HSM algorithm, respectively.  Modelled estimates of the complexity for the $\approx 6.1$ million km$^2$ Amazon River basin show a reduction from $6.7$ billion to $1.4$ million reads.

\end{abstract}


 \end{frontmatter}



\section{Introduction}
Watersheds delineation is the ``basic modelling element" Tesfa et al 2011  \cite{tesfa} for hydrological problems.  Watershed are important for a number of areas of interest including biological \cite{daniel_watershed_2011,wang_influences_1997}, ecological \cite{malvadkar_comparison_2015, daniel_watershed_2011}, infrastructure (e.g.cadastral) flood \cite{al-sabhan_real-time_2003}, and engineering.  A number of software packages (e.g. TAUDEM \cite{taudem}, ESRI's hydrologic analysis tools \cite{esrihtools}) have been designed to support the creation of watersheds as polygonal boundaries (e.g., a vector of coordinates that represent a polygon on a two dimensional planar surface).  

In this manuscript we focus on creating watershed boundaries from digital flow direction grids.  Baker et al 2006. \cite{baker} discussed standard practices to create polygonal watershed boundaries using a flow direction grid approach as 1) creating a digital elevation model (DEM), 2) filling spurious sinks, 3) creating a flow direction grid, 4) burning in fluvial hydrological features (e.g., streams and rivers),  and 5) using an accumulation of cell contributions across the landscape.  Our proposed algorithm substitutes for the last step a  general data structure and its corresponding labelling model called the modified nested set as described in Haag and Shokoufandeh 2017 \cite{haag}.  A novel algorithm Haag Shokoufandeh Marching (HSM) is then applied to retrieve watershed boundaries from the modified nested set data model.

The nested set model as described by Celko 2012 \cite{celko_joe_2012} and Haag and Shokoufandeh 2017 \cite{haag} is a modified depth first graph traversal algorithm.  Starting at the root or the pour point in a $D8$ grid, the flow direction grid is traversed and for each cell the discovery time, finish time, and distance from the root are recorded.  Refer to Haag and Shokoufandeh 2017 \cite{haag} for a detailed description and algorithms to implement and run the modified nested set model within a flow direction grid or a vector of connected catchments such as the NHD plus Version 2~\cite{david_river_2011}.

This manuscript describes a new technique to retrieve watershed boundaries from a D8 Flow direction grid.  The HSM algorithm's retrieval complexity scales directly from the number of vertices necessary to delineate a watershed.  Existing algorithms retrieval complexity scales in relation to the number of grid cells that make up the interior of the watershed boundary.  Therefore the HSM algorithm outperforms existing techniques geometrically depending on the shape of the watershed.


\section{Notations and Data}
\subsection{Notation}
\label{subsec1}

\begin{figure}[t!]
    \centering
    \includegraphics[scale =0.7]{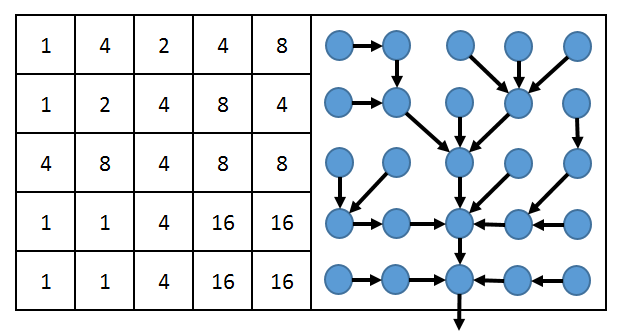}
    \caption{Flow Direction Grid and associated graph $G(V,E)$ representation.}
 \label{FlowGrid}
 \end{figure}

Before describing the HSM algorithm in detail, we discuss the notation and  structures used throughout the rest of the manuscript. We use the set of natural numbers $\{0,1,2,3\}$ for encoding cardinal directions North, East, South, and West, respectively. We apply the HSM algorithm to delineate watershed boundaries for any surface using a regularly spaced flow direction $D8$ grid~\cite{jenson} as input.  Within a $D8$ structure every grid cell contains a value in $\{2^0,2^1, ... , 2^7\}$ that denotes connectivity to one of its eight neighboring grid cells.  We treat this flow grid as a graph (Figure~\ref{FlowGrid}) $G=(V,E)$ whose vertex set $V = \{v_1, v_2, ... , v_n \}$ is congruent to grid cells and its directed edge set  $E \subset \{\overrightarrow{(v_i,v_j)}|\ 1\le i,j \le n\}$ is congruent to the flow direction values of individual vertices. Specifically, each node $u\in V$ has exactly one $D8$ corresponding to an outgoing flow to a cell $v$ which will result in  exactly one edge $e = \overrightarrow{(u,v)}$ in the edge-set $E$.  Additionally, since each node in $v\in V$ corresponds to a regular grid cell, we can describe its location using index pair $\rho(v) = (x(v),y(v))$.\\

\begin{figure}[b!] 
    \centering
    \includegraphics[scale = .35]{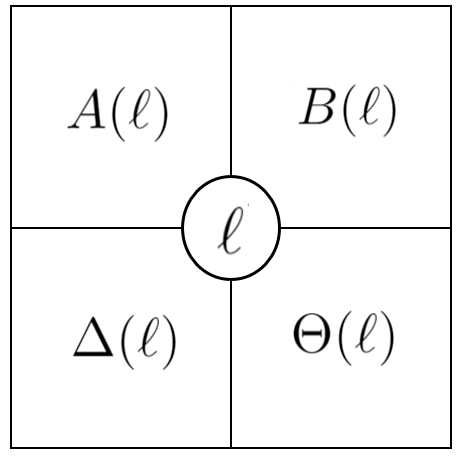}
    \caption{Labels for neighboring cells of a lattice location $\ell \in {\cal L}$ denoted as  \{${\cal A}(\ell),{\cal B}(\ell), \Omega(\ell), \Delta(\ell)$\}.  }
\label{CellLabels}
\end{figure}

We create a secondary abstraction of the $D8$ flow direction grid to denote the lattice corners of each grid cell $v \in V$. Specifically, we use ${\cal L}=(\Lambda , \Gamma)$ to denote a regular square lattice, with the set $ \Lambda \subset \R^2$ denoting the set of lattice points, and  $\Gamma \subset \R^2 \times \R^2$ as the set of edges connecting lattice points. Similar to vertices of graph $G$, each lattice point $\ell$ in  $\Lambda$ is uniquely identified by an index pair $(x(\ell),y(\ell))$. We assume without loss of generality that ${\cal L}$ is a periodic lattice.  A $k$-step lattice walk from lattice point $p_0$ to lattice point $p_k$ is a sequence $W = \{p_0, ..., p_k\}$ of elements in $\Lambda$ such that $\forall i \in \{0, ..., k-1 \}$ ; $(p_i, p_{i+1}) \in \Gamma $.  We define ${\cal F}$ as the set of faces or cells of ${\cal L}$ and note that there is an equivalence between the vertex set $V$ of graph $G$ and the set of  faces ${\cal F}$ for lattice ${\cal L}$. We note that the boundary of any watershed for lattice ${\cal L}$ is a closed lattice walk (lattice cycle), i.e., $p_0 = p_k$. Given a lattice point $\ell \in {\cal L}$, we use $A(\ell ),  B(\ell),  \Theta (\ell),$ and $\Delta (\ell)$, if they exist,  to donate its four neighboring lattice faces in $\cal F$, i.e.,  vertices in $G$ (see Figure~\ref{CellLabels}). Conversely, for a face $u \in {\cal F}$, we use the $\alpha (u), \beta(u), \delta (u), \gamma(u)$ to denote its four boundary lattice points (see Figure~\ref{LatticeDirection}).

\begin{figure}[t!] 
    \centering
    \includegraphics[scale = .65]{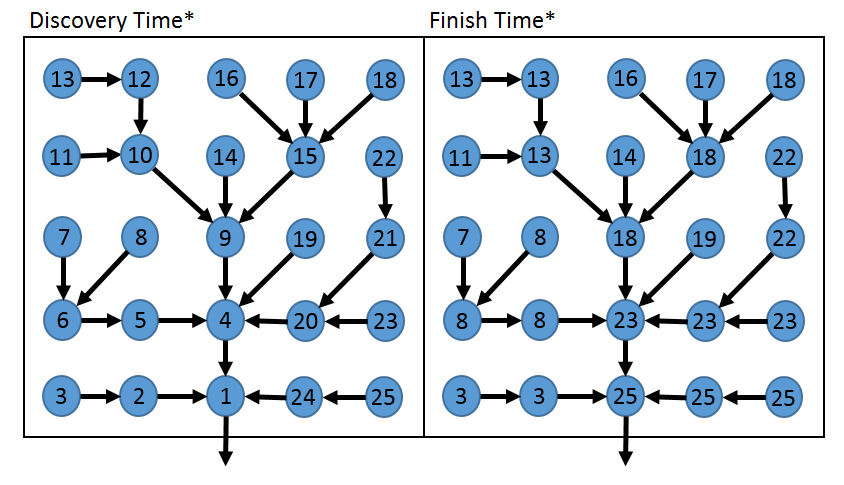}
    \caption{Graph labelled with discovery and finish times using the Modified Nested Set Algorithm.}
\label{HSNLables}
\end{figure}

As a pre-processing step prior to running the HMS algorithm we apply the MNS algorithm to the $D8$ grid as described in (Haag in Shokoufandeh 2017)~\cite{haag}.  This algorithm will associate two additional attributes of discovery time $d(v)$ and finish time $f(v)$ to each node $v\in V$.  The attribute $d()$ is unique for every grid cell, while multiple grid cells can receive the same $f()$ value.  Figure~\ref{HSNLables} illustrates discovery and finish times for a sample graph. Finally, the marching algorithm takes as input any cell $v^*\in V$ and returns its watershed boundary, in the form of a closed lattice walk $\Omega(v^*)$, made up of a list of lattice points in $\cal L$.  

\subsection{Source Data}
\label{subsec2}
\begin{figure}[t!]
    \centering
    \includegraphics[scale = .7]{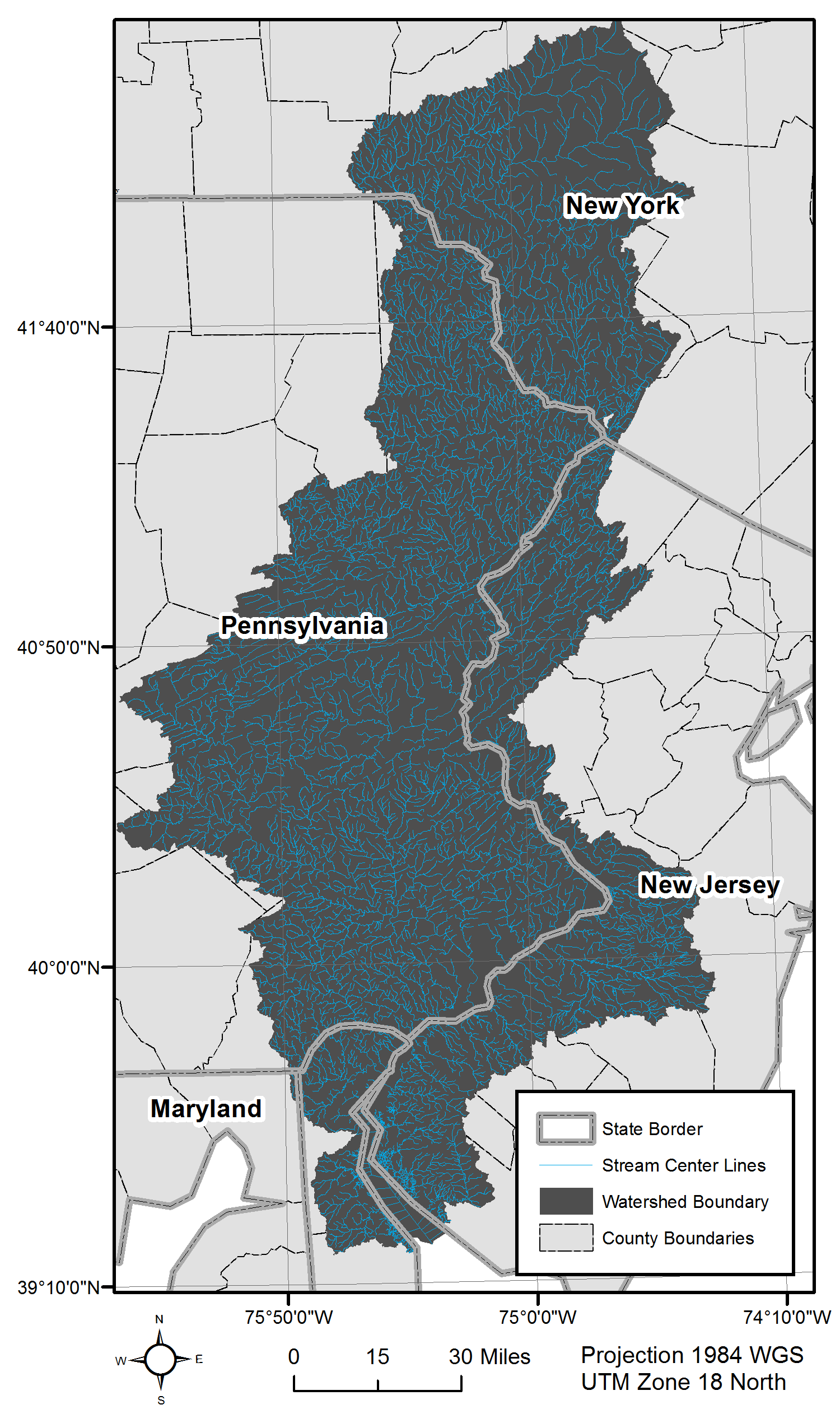}
    \caption{Overview Map of Delaware River Watershed Study Area.}
 \label{DRWS}
\end{figure}

We applied the marching algorithm to the Delaware River Watershed using the Flow Direction Grid provided as part of the National Hydrography Plus Version 2 database (http://www.horizon-systems.com/NHDPlus/NHDPlusV2\_02.php).  This dataset was downloaded for the mid-Atlantic version region 02-A  (EPA and Horizon Systems).  The DEM was a subset based on the watershed boundary for the Delaware River Watershed with the root node (hydro sequence 200005438) and all stream reaches where the TerminalPa $= 200005438$ or the confluence of the Delaware River with the Delaware Bay  $-75.35717$, $39.2753$. This is the same study area used in our previous work (Haag and Shokoufandeh 2017)~\cite{haag}. This resulted in a study area of $\approx 36,000\ {\rm km}^2$ (Figure~\ref{DRWS}).

\section{Method}
Our algorithm relies on certain structural properties of a large flow direction raster grid. Specifically, we distinguish between {\em root} cells and {\em internal} cells of a flow grid. The root cells are identified as grid cells that flow into cells with null values (e.g. large bodies of water, oceans, and lakes), or off the grid.  All other grid cells that flow through to cells with valid flow directions are internal cells. As we discussed earlier, our marching algorithm utilizes an auxiliary data structure known as MNS that captures the hierarchy and nested structure of watersheds. The MNS data structure encodes the hierarchical complexity of a watershed using a pair of attributes, $d(v)$ and $f(v)$,  for each cell $v$ in flow grid $G$. 

The marching algorithm receives as its input a vertex $v^*$ as its staring cell from the flow direction grid $G$ and in incremental fashion builds the boundary of a complete vector polygon representing the watershed $\Omega(v^*)$ of all grid cells that drain through pour point $v^*$. The polygonal chain $\Omega(v^*)$ is a closed lattice walk defined on ${\cal L}$, the dual lattice for flow grid $G$.  The correctness of our incremental algorithm relies on the observations that every  boundary grid cell of  of $\Omega(v^*)$, with a valid flow direction value is guaranteed to have at least one adjacent neighbor that is not in its watershed that the cell it flows into. For the initial cell $v^*$, the incremental algorithm is guaranteed to find adjacent lattice points belonging to $\Omega(v^*)$ (Proposition~\ref{prop1}). 
\begin{proposition}
\label{prop1}
Given the initial grid vertex $v^*$ with a valid flow direction in $\{2^0, 2^1, ... , 2^7\}$, at least one of its 4 adjacent lattice points will be coincident on the polygon boundary representing $\Omega(v^*)$.
\end{proposition}

\begin{figure}[t!] 
    \centering
    \includegraphics[scale = 0.75]{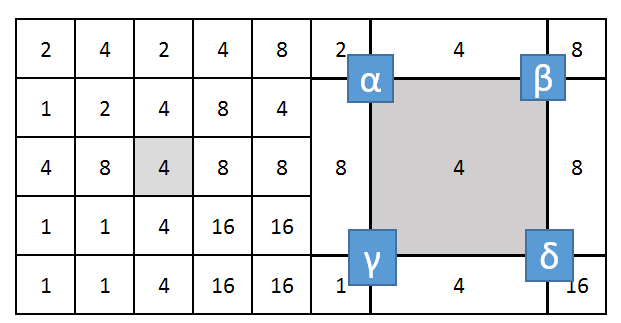}
    \caption{Lattice imposed on flow direction grid.}
\label{LatticeDirection}
\end{figure}

This in turn allows us to efficiently identify boundary lattice points and extend the chain $\Omega(v^*)$ (Proposition~\ref{prop2}). Specifically, starting at grid cell $v^*$, its outgoing edge $D8(v^*)$ will be used to identify the lattice point(s) that share a boundary with $v^*$.  Note that  if  $D8(v^*)$ belongs to set $\{1,4,16,64\}$, the cell $v^*$ and the destination cell of the edge $D8(v^*)$ share two lattice points,  while for $D8(v^*)$ in $\{2,8,32,128\}$ share a single lattice point (Figure~\ref{LatticeDirection}). At each step, the algorithm identifies a new lattice point to extend the march (chain $\Omega(v^*)$) based on the direction of the march (clockwise or counterclockwise) and the flow direction $D8$ attribute of the starting grid cell.  
\begin{proposition}
\label{prop2}
For any lattice location on the boundary of $\Omega(v^*)$ at least two of its neighboring 4 lattice locations are  coincident on the polygon boundary representing $\Omega(v^*)$.
\end{proposition}

While we have chosen to use a consistent counterclockwise probe as the reference direction in every step, the algorithm is symmetric and would perform similarly irrespective of the consistent reference direction of incremental steps.  Proposition~\ref{prop3} is a consequence of this consistent walk.  
\begin{proposition}
\label{prop3}
Assuming the algorithm follows  a counterclockwise march around the lattice the interior of the polygon $\Omega(v^*)$ can always be maintained on the left hand side of the boundary lattice points at each step of the algorithm.
\end{proposition}

After identifying a starting lattice point, the march will continue by examining the immediate counterclockwise lattice point with respect to the previous move. For example, if the march starts by moving to the top right and the the HSM march is moving in the North direction (cardinal direction with value 0), the next boundary check will occur to the East (value 1).  At every iteration of the HSM algorithm, we identify a potential new location (lattice point) for extending the march using the current lattice location and the march direction using the {\bf LatticeMove} function. Next, the identified lattice location is examined for membership in $\Omega (v^*)$ using a Boolean function, referred to as  {\bf BoundaryPoint}. Specifically, the new lattice point is on the boundary if and only if at least one of its four neighboring cells has its $d()$ values between $d(v^*)$ and $f(^*)$, and at least one of its four neighboring cells has its $d()$ value outside the range $d(v^*)$ and $f(v^*)$. If the candidate lattice location violates the aforementioned condition, it is not part of $\Omega(v^*)$, and the HSM algorithm will proceed to the next lattice point in a counterclockwise order as  potential march direction. In the case of the previous example, it would search through North, West, and South until a boundary lattice point is identified.  If the {\bf BoundaryPoint} function returns a  {\em true} value then the candidate lattice point will be added to the watershed boundary, the search direction is moved one position on a counterclockwise rotation. This structural property of $\Omega (v^*)$ is the subject of Proposition~\ref{prop4}.

\begin{proposition}
\label{prop4}
The marching algorithms will advance in each iteration of the algorithm to a lattice point which is neither inside nor outside of the watershed, i.e., it is on the boundary of  $\Omega (v^*)$. Moreover, the advance step can be carried in constant time. 
\end{proposition}

\begin{figure}[b!] 
    \centering
    \includegraphics[scale = 0.825]{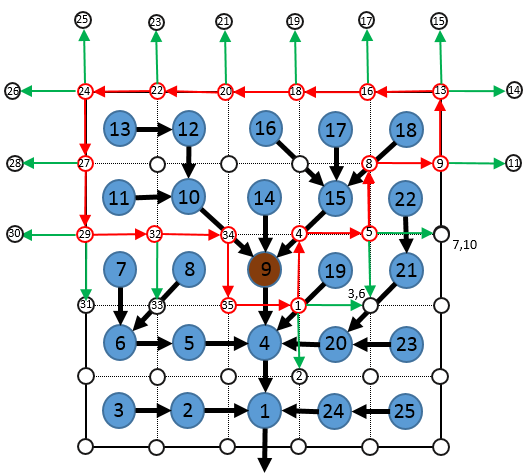}
    \caption{Marching Algorithm marching steps for $v^*$ with $d(v^*)=9$ ($f(v^*) =18$). Edges in red indicate successful march, while green edges show negative results for boundary condition.  Graph vertices (blue circles) are labelled with the MNS algorithm discover time. HSM lattice locations (small black circles) are labelled with the recursive step in the HSM algorithm. }
 \label{HSMAlg}
\end{figure}

In the worst case during each iteration of the march, the algorithm will examine three lattice values (otherwise it would fold upon itself returning along the same edge it came in) and proceeds to identify the next lattice point that is on the boundary of $\Omega(v^*)$. The algorithm terminates when it returns to the first lattice edge crossed during the march identifying a closed polygon. Figure~\ref{HSMAlg} illustrates the result of executing the algorithm on an example graph (starting at node $v^*= 9$). Proposition~\ref{prop5} formalizes the computational complexity of the marching algorithm. 

\begin{proposition}
\label{prop5}
Every lattice point belonging to $\Omega(v^*)$ will be visited at most twice. This results in a $O(|\Omega(v^*)|)$ complexity for the marching algorithm. 
\end{proposition}

\newpage
 \textbf{Function 1. LatticeMove ($\ell,\mu$) }  
 \begin{footnotesize}
 \begin{algorithmic}[1]
 \State {\bf Input:} Lattice point $\ell \in {\cal L}$ and $\mu$ direction $\in$ \{1,2,3,4\}.  Here $\ell$ is uniquely identified as $\ell=\Lambda(x(\ell), y(\ell))$,
\State {\bf Output:} new lattice cell in direction $\mu$.\\
${\cal P} = (\mu + 1) \pmod 2$ \\
${\cal S} = \lceil \mu / 2\rceil$ \\
$\delta x =  (-1)^{(1+{\cal S})} |{\cal P}|$ \\
$\delta y = (-1)^{(1+{\cal S})}|{\cal P} -1|$ \\
\Return $\Lambda(x(\ell)+\delta x,y(\ell)+\delta y)$
\end{algorithmic} 

\end{footnotesize} 

\textbf{Function 2. BoundaryPoint($\ell, v^*$)}  
 \begin{footnotesize}

 \begin{algorithmic}[1]
 \State {\bf Input:}  $\ell$ lattice cell $\in {\cal L}$  and $v^*$ starting vertex $\in G$:
 \State {\bf Output:} A Boolean indicating $\ell \in \Omega_{v^*}$:
 \State {\bf value = False } 
\If{$d(A(\ell) ) \in [d(v^*),f(v^*)]$ or \\ 
\ \ \ $d(B(\ell) ) \in [d(v^*),f(v^*)]$ or \\ 
\ \ \ $d(\Theta(\ell) ) \in [d(v^*),f(v^*)]$ or \\ 
\ \ \ $d(\Delta(\ell) ) \in [d(v^*),f(v^*)]$} 
\If{ $d(A(\ell) ) \notin [d(v^*),f(v^*)]$ or \\ 
 \ \ \ \ \ \ \ \ \  $d(B(\ell) ) \notin [d(v^*),f(v^*)]$ or \\ 
 \ \ \ \ \ \ \ \ \  $d(\Theta(\ell) ) \notin [d(v^*),f(v^*)]$ or \\ 
 \ \ \ \ \ \ \ \ \  $d(\Delta(\ell) ) \notin [d(v^*),f(v^*)]$} 
\State {\bf value = True } 
\EndIf{}
\EndIf{} \\
\Return {\bf value}
\end{algorithmic} 
\end{footnotesize} 

\textbf{Algorithm 1. Watershed Marching Algorithm ($G,{\cal L},v^*$)} 
\begin{footnotesize}
\begin{algorithmic}[1]
\State {\bf Input:} Graph $G(V,E)$ such that $\forall v \in V$ and Lattice ${\cal L}$ such that $\forall$ $\ell \in {\cal L}$: \\
\ \ \ \ \  - $v$ has discovery, $d(v)$, and finish, $f(v)$, values computed by MNS algorithm. \\
\ \ \ \ \  - Lattice point $\ell \in {\cal L}$ has unique grid coordinates pair $\Lambda(x(\ell), y(\ell))$,\\
\State {\bf Output:} Watershed as a (cyclic)  array of lattice points:\\ 
\ \ \ \ \ \ \ \ \ \ \ \ \ \  $\Omega(v^*) = \left <\Lambda({x_1,y_1}),\Lambda({x_2,y_2}), \ldots ,\Lambda({x_k,y_k}),\Lambda({x_1,y_1}), \Lambda({x_2,y_2}) \right >$. 
\State ${\cal C}= \left <\beta(),\gamma(),\gamma(),\delta(),\delta(),\alpha(),\alpha(),\beta()\right >$
\Comment Aux. array of possible lattice functions 
\State ${\cal D}= \left <1,1,2,2,3,3,0,0 \right>$
\Comment Aux. array of possible cardinal transition directions 
\State $idx = \lg D8(v^*)$   
\Comment Aux. var for the index in arrays  ${\cal C}$ and ${\cal D}$
\State $l_{v^*} = {\cal C}[idx](\rho(v^*))$
\Comment Select the corner function           \Comment Identify Starting Lattice location
\State $\Omega(v^*)=\left <  l_{v^*}\right >$  \Comment Store the lattice coordinate pair in the final watershed
\State $\mu \leftarrow {\cal D}[idx]$          \Comment Store march direction state

\While {($\Omega[1] \neq \Omega[|\Omega|-1]$  and $\Omega[2] \neq \Omega[|\Omega|])$ }
\Comment Stop when the first edge is recrossed.
\State $\ell\leftarrow {\bf LatticeMove}(\Omega[|\Omega|], \mu).$ \Comment apply march function based on direction

\If {  {\bf BoundaryPoint($\ell , v^*$)} } 

\State $\Omega \leftarrow \Omega \uplus \Lambda ( x(\ell),y(\ell) ) $  \Comment append new coordinates to watershed boundary
\State $\mu \leftarrow ( \mu +1) \mod 4$ \Comment Advance the current vertex. 

\Else 
\State $\mu \leftarrow (\mu -1) \mod 4$ \Comment Set next direction Counterclockwise.
\EndIf{}
\EndWhile
\State $\Omega \leftarrow \Omega[1,...,|\Omega| - 1]$ \Comment remove the duplicate vertex at the end of the polygon.
\State \Return $\Omega$
\end{algorithmic} 
\end{footnotesize}

\section{Results}

\subsection{Data model pre-processing cost}
\label{subsec5.2}

The cost to convert the original flow direction grid into the modified nested set model is linear based on the size of the grid (Haag and Shokoufandeh 2017) \cite{haag}. The algorithm crosses every edge between grid cells exactly two times on the depth first graph traversal. The $D8$ flow direction grid only allows one edge (the downstream connection) for every grid cell (except for the root which has no edges), therefore the application runs in exactly $(2n-1)$ where $n$ is the number of grid cells or polygonal catchments. The data structures $G(V,E)$ and lattice ${\cal L}$ can both be derived from outputs of the modified nested set algorithm and therefore require no further processing or storage resources.

\subsection{Data model storage size}

We compare the size of the data model required to support HSM algorithm to the initial size of the input $D8$ flow direction grid.  Because the algorithm relies on the original data to initiate the march, a reduction in size is not possible. Conversely, because the attributes (discovery $d()$, finish $f()$, and $x$ and $y$ coordinates) are stored for every cell location the size of the final data set is linear and constant with respect to the size of the initial grid cell.

\subsection{Comparison to existing watershed delineation algorithms} 
The proposed HSM algorithm was compared to existing watershed creation algorithms for the 14,718 NHDPlus V2 \cite{david_river_2011} catchments within the Delaware River Watershed.  For each catchment the total query complexity for the generic grid processing algorithms (e.g. Gage Watershed from Taudem~\cite{taudem} or ESRI's watershed function~\cite{esrihtools}) was compared to the proposed HSM algorithm.  Specifically, the number of  $30m$ grid cells that was required to process to create $\Omega(v^*)$ was compared between the two approaches.  We did not include in this comparison pre-processing steps but rather we assume as inputs to the original algorithm a standard flow direction $D8$ grid and to the HSM algorithm the discovery and finish time grids from Haag and Shokoufandeh 2017~\cite{haag}.  We believe that this is a valid approach because the conversion from raw DEM data to the sink filed $D8$ flow direction model requires multiple sweeps across the raw graph.  The conversion to the discovery and finish time grid as required for this algorithm require two more sweeps across the entire grid and therefore this preprocessing step will run in linear constant time in relation to existing steps. 

Figure~\ref{Complexity} illustrates the results for the comparison between the $14,718$ catchments.  The standard grid processing algorithm requires between $1$ and $\approx 34$ million while the proposed algorithm requires between $1$ and $\approx 44$ thousand vertices (grid cells) to be processed.  We fit a linear regression least squares model against the logarithm of $y = c x ^{b}$ ($\ln y = \ln c + b \ln x$) resulting in $y = 0.1967x^{1.7986}$ with an $r^2$ value of $0.98$.  If we extend this equation to larger watersheds, for example the $\approx 6.1$ million km$^2$ Amazon River, it would require $\approx 6.8$ billion $30$ meter grids cells while HSM would require $\approx 1.4$ million marches. This is a reduction of $99.97\%$ read operations vs. known algorithms. 

As the horizontal grid resolution of DEM decrease the proposed algorithms will outpace traditional techniques at a faster rate.  For example the 3 Digital Elevation Program (3DEP) being lead by the United States Geological Service (USGS) already provides 1 meter DEM datasets for select locations within the coterminous United States \cite{3dep}.  When this resolution data becomes available for the Amazon river basin it will turn the original $6.1$ billion grid necessary to delineate the Amazon basin into a $5.5$ trillion grid.  This is caused by the fact that DEMs increase in resolution geometrically (e.g. the difference in size between 30 meter DEM and 1 meter DEM is $(30/1)^2 ) = 900$.  Applying our linear regression model Figure\ref{Complexity} we estimate that the proposed algorithm would need to visit roughly $28$ million grid cells.  The higher the DEM resolution the larger the advantage the HSM algorithm has compared to existing data structures.

\section{Discussion}
\subsection{Complexity reduction}
\begin{figure}[t!] 
    \centering
    \includegraphics[scale = .8]{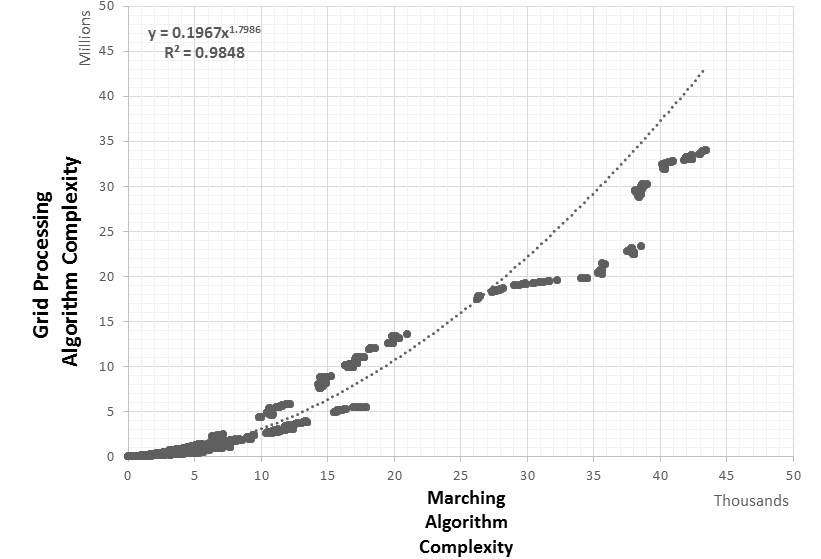}
    \caption{An illustration of the complexity comparison between the proposed algorithm vs. existing techniques for $n$ = 14,718 watershed in the Delaware River Drainage}
\label{Complexity}
\end{figure}

The observed reduction in computational complexity is consistent with the observation that the proposed algorithm has the same complexity as the watershed boundary or to put it another way in the length of the polygonal chain $\Omega(v^*)$. In contract, existing algorithms scale in proportionality to the area of the watershed.  We expect that based on the watershed shape the proposed algorithm retrieval complexity would be geometrically smaller then existing techniques.  Empirical tests conducted within the Delaware River Watershed show a geometric reduction in processing time Figure\ref{Complexity}.  When we extend the empirical results to other watershed we predict considerable, $\approx{} 99.9\%$, reduction in computational complexity.  Additionally, the advantage that the HSM algorithm has vs existing techniques will become more prominent as DEMs continue to increase in resolution. For example select parts of the Conterminous United States already have 1 meter DEMs as part of the USGS Three Dimensional Elevation Program~\cite{3dep}.  This will result in a data-set with $900$ times the number of nodes that were used in this analysis.

\subsection{Potential issues with this data model}
In this section we discuss a number of potential issues that could arise by using the proposed data model vs. existing techniques.  The first involves endoheric basins (e.g. the Caspian Sea or the Great Salt lake Watershed).  In the rare situation where endoheric basins are completely surrounded by an exorheic basin(s) the HSM algorithm would subsume the internal watershed.  To efficiently solve this problem a standard spatial intersection with the watershed produced by the HSM algorithm vs. the internal centroid for endoheric basins will determine if a watershed contains internal basins.  If a centroid for an endoheric basin is contained within the HSM boundaries then the boundary of the internal endoheric basin can be appended to the HSM algorithm as an internal boundary (no spatial processing necessary). 

The second potential issue arises within watersheds where stream flow is divergent (e.g., oxbows or tidal areas).  This issue is discussed in more detail in Haag and Shokoufandeh 2017 \cite{haag}.  We note that assuming water flow is one directional is common among watershed and stream routing models and in fact is an underlying structural assumption for the $D8$ flow direction grid.

\newpage{}
\bibliographystyle{acm}
\bibliography{fromzotero}

\newpage{}


\end{document}